\documentclass[12pt,a4paper]{article}

\usepackage[utf8]{inputenc}
\usepackage[T1]{fontenc}
\usepackage{amsmath,amssymb}
\usepackage{booktabs}
\usepackage{graphicx}
\usepackage{hyperref}
\usepackage{multirow}
\usepackage{xcolor}
\usepackage{tikz}
\usetikzlibrary{shapes.geometric, arrows.meta, positioning, fit, backgrounds}
\usepackage{natbib}
\usepackage[margin=2.5cm]{geometry}
\usepackage{setspace}
\usepackage{authblk}

\begin{document}

\title{Fine-Tuning Large Language Models for Automatic Detection of Sexually Explicit Content in Spanish-Language Song Lyrics}

\author[1]{Dolores Zamacola S\'anchez de Lamadrid}
\author[1,2]{Eduardo C. Garrido-Merch\'an\thanks{Corresponding author. E-mail: ecgarrido@comillas.edu}}
\affil[1]{Universidad Pontificia Comillas, ICADE, Madrid, Spain}
\affil[2]{Instituto de Investigaci\'on Tecnol\'ogica (IIT), Madrid, Spain}

\date{}

\maketitle

\begin{abstract}
The proliferation of sexually explicit content in popular music genres such as reggaeton and trap, consumed predominantly by young audiences, has raised significant societal concern regarding the exposure of minors to potentially harmful lyrical material. This paper presents an approach to the automatic detection of sexually explicit content in Spanish-language song lyrics by fine-tuning a Generative Pre-trained Transformer (GPT) model on a curated corpus of 100 songs, evenly divided between expert-labeled explicit and non-explicit categories. The proposed methodology leverages transfer learning to adapt the pre-trained model to the idiosyncratic linguistic features of urban Latin music, including slang, metaphors, and culturally specific double entendres that evade conventional dictionary-based filtering systems. Experimental evaluation on held-out test sets demonstrates that the fine-tuned model achieves 87\% accuracy, 100\% precision, and 100\% specificity after a feedback-driven refinement loop, substantially outperforming both its pre-feedback configuration and a non-customized baseline ChatGPT model. A comparative analysis reveals that the fine-tuned model agrees with expert human classification in 59.2\% of cases versus 55.1\% for the standard model, confirming that domain-specific adaptation enhances sensitivity to implicit and culturally embedded sexual references. These findings support the viability of deploying fine-tuned large language models as automated content moderation tools on music streaming platforms. Building on these technical results, the paper develops a public policy proposal for a multi-tier age-based content rating system for music---analogous to the PEGI system for video games---analyzed through the PESTEL framework and Kingdon's Multiple Streams Framework, establishing both the technological feasibility and the policy pathway for systematic music content regulation.
\end{abstract}


\vspace{1em}


\section{Introduction}
\label{sec:introduction}

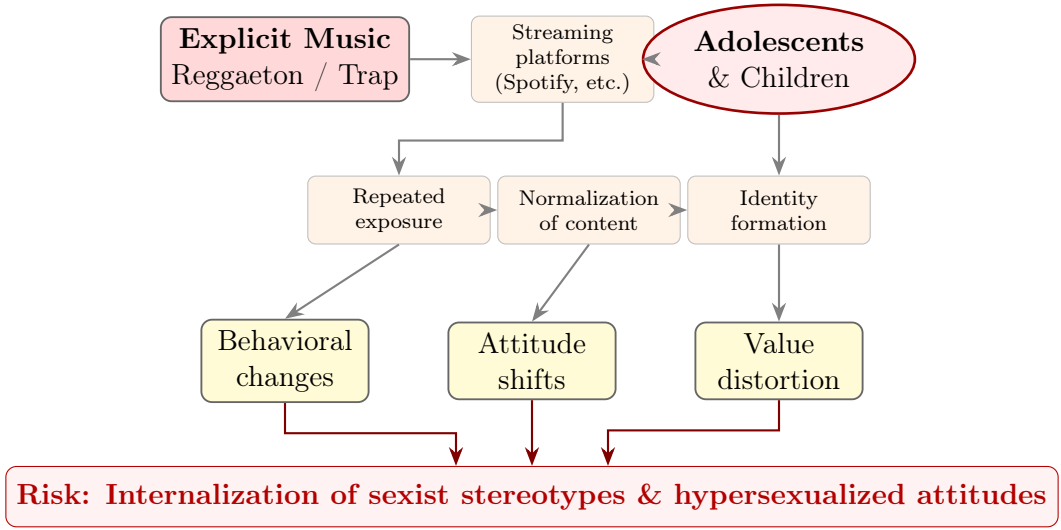
\begin{figure}[t]
\centering
\begin{tikzpicture}[
    source/.style={rectangle, rounded corners=4pt, minimum width=2.2cm, minimum height=1cm,
        align=center, font=\small, draw=black!60, fill=red!15, line width=0.7pt},
    pathway/.style={rectangle, rounded corners=3pt, minimum width=2.4cm, minimum height=0.9cm,
        align=center, font=\scriptsize, draw=gray!50, fill=orange!10, inner sep=3pt},
    impact/.style={rectangle, rounded corners=4pt, minimum width=2.2cm, minimum height=1cm,
        align=center, font=\small, draw=black!60, fill=yellow!20, line width=0.7pt},
    vulnerable/.style={ellipse, minimum width=2.8cm, minimum height=1.4cm,
        align=center, font=\small, draw=red!60!black, fill=red!8, line width=1pt},
    arrow/.style={-{Stealth[length=2.5mm]}, thick, draw=black!50}
]

\node[source] (music) at (0, 3) {\textbf{Explicit Music}\\Reggaeton / Trap};

\node[pathway, right=0.8cm of music] (stream) {Streaming\\platforms\\(Spotify, etc.)};

\node[vulnerable] (youth) at (6.5, 3) {\textbf{Adolescents}\\\& Children};

\node[pathway] (repeat) at (1.5, 1) {Repeated\\exposure};
\node[pathway] (normal) at (4, 1) {Normalization\\of content};
\node[pathway] (ident) at (6.5, 1) {Identity\\formation};

\node[impact] (behav) at (0, -1) {Behavioral\\changes};
\node[impact] (attit) at (3.25, -1) {Attitude\\shifts};
\node[impact] (values) at (6.5, -1) {Value\\distortion};

\node[rectangle, rounded corners=4pt, draw=red!70!black, fill=red!5,
      minimum width=8cm, minimum height=0.8cm, align=center, font=\small\bfseries,
      text=red!70!black] (warning) at (3.25, -2.8)
      {Risk: Internalization of sexist stereotypes \& hypersexualized attitudes};

\draw[arrow] (music.east) -- (stream.west);
\draw[arrow] (stream.east) -- (youth.west);
\draw[arrow] (youth.south) -- (ident.north);
\draw[arrow] (stream.south) -- ++(0, -0.5) -| (repeat.north);
\draw[arrow] (repeat.east) -- (normal.west);
\draw[arrow] (normal.east) -- (ident.west);

\draw[arrow] (repeat.south) -- (behav.north);
\draw[arrow] (normal.south) -- (attit.north);
\draw[arrow] (ident.south) -- (values.north);

\draw[arrow, draw=red!50!black] (behav.south) -- ++(0, -0.4) -| ([xshift=-1cm]warning.north);
\draw[arrow, draw=red!50!black] (attit.south) -- (warning.north);
\draw[arrow, draw=red!50!black] (values.south) -- ++(0, -0.4) -| ([xshift=1cm]warning.north);

\end{tikzpicture}
\caption{Pathways through which sexually explicit content in music can influence adolescents and children. Repeated exposure through streaming platforms leads to normalization of explicit content during critical periods of identity formation, potentially resulting in behavioral changes, attitude shifts, and value distortion. Research has documented associations between exposure to degrading sexual content in music and shifts in adolescent sexual behavior \citep{Martino2006}.}
\label{fig:social_impact}
\end{figure}

Music has long been recognized as one of the most universal forms of artistic expression, with the capacity to shape emotional states, influence behavior, and transmit cultural values across generations. While this communicative power is often celebrated, it also carries the potential for harm when lyrical content normalizes violence, misogyny, or hypersexualized discourse, particularly among impressionable listeners. In recent years, the global ascendancy of urban Latin genres---most notably reggaeton and trap---has intensified public debate around this issue. Bad Bunny, one of the most prominent reggaeton artists, was the most-streamed artist on Spotify for three consecutive years from 2020 to 2022 \citep{Spotify2022}, underscoring the massive reach of a genre whose lyrics frequently contain sexually explicit material. The psychological literature has established that sustained exposure to degrading sexual content in music lyrics is associated with shifts in adolescent sexual behavior and attitudes \citep{Martino2006}, and educational researchers have documented that reggaeton lyrics often reproduce sexist stereotypes and normalize the objectification of women \citep{DiezGutierrez2023}. These concerns are compounded by the developmental vulnerability of children and adolescents, who are in a formative stage of identity construction and are therefore more susceptible to internalizing the messages embedded in the music they consume. Figure~\ref{fig:social_impact} illustrates the pathways through which explicit musical content can influence young listeners.

The need for automated systems capable of identifying and flagging explicit content in song lyrics is thus both socially urgent and technically challenging. The Recording Industry Association of America (RIAA) introduced the voluntary ``Parental Advisory'' label in the 1990s as a self-regulatory measure \citep{RIAA2021}, yet this system depends on manual, inconsistent, and voluntary application by record labels, leaving significant gaps in coverage. Analogous content advisory systems have become standard in film and television streaming platforms such as Netflix and HBO, which provide granular content warnings prior to playback, but the music industry has lagged considerably in implementing comparable automated safeguards. A robust, scalable system for detecting explicit lyrical content could be integrated into music streaming platforms such as Spotify and Apple Music, enabling not only advisory labels but also automated parental controls, age-appropriate playlist curation, and content-aware recommendation systems.

The technical difficulty of this task stems from the nature of explicit content in urban Latin music. Unlike straightforward profanity detection, which can be addressed through keyword blacklists, the identification of sexually explicit material in reggaeton and trap lyrics requires deep semantic understanding. These genres employ an elaborate vocabulary of slang, metaphors, double entendres, and culturally specific euphemisms---expressions such as ``bellaquear,'' ``perreo,'' and ``mamacita''---that carry strong sexual connotations but are absent from standard dictionaries and would be missed entirely by rule-based filtering systems. Early work by \citet{Chin2018} demonstrated that dictionary-based filtering achieves only 61\% F1-score on explicit lyrics detection, precisely because of its inability to capture contextual meaning. Machine learning approaches using bag-of-words representations and decision trees improved performance to 78\% F1-score on Korean-language lyrics \citep{Chin2018}, but these methods still operate at the surface level of individual tokens rather than grasping the semantic relationships that determine whether a passage is explicitly sexual, romantically suggestive, or entirely benign.

The emergence of transformer-based large language models (LLMs) has fundamentally altered the landscape of natural language processing (NLP), offering architectures capable of capturing long-range dependencies and contextual nuance that were previously inaccessible to statistical and shallow neural approaches. The seminal work of \citet{Vaswani2017} introduced the self-attention mechanism, which allows every token in a sequence to attend to every other token, thereby enabling the model to weigh the relevance of distant words when constructing representations. The Generative Pre-trained Transformer (GPT) family of models, developed by OpenAI and built upon the decoder-only variant of the transformer architecture, has demonstrated remarkable versatility across tasks ranging from text generation and sentiment analysis to semantic classification \citep{Brown2020}. The key insight behind transfer learning with LLMs is that a model pre-trained on vast corpora of general text acquires rich statistical representations of language that can subsequently be adapted to specialized downstream tasks through fine-tuning on comparatively small domain-specific datasets.

This paper presents a study on the application of fine-tuned GPT models to the automatic detection of sexually explicit content in Spanish-language reggaeton and trap lyrics. A corpus of 100 songs was compiled, with 50 labeled as explicit and 50 as non-explicit by a domain expert, and a reference table of explicit phrases was constructed to guide the model toward recognizing not only overt sexual language but also implicit references, metaphors, and genre-specific slang. The model was fine-tuned using supervised learning with binary cross-entropy loss and the AdamW optimizer, and its performance was evaluated through standard binary classification metrics derived from confusion matrices. A feedback-driven refinement loop was then applied to address classification errors from the initial evaluation, and a comparative analysis was conducted against a non-customized ChatGPT baseline to quantify the value added by domain-specific fine-tuning.

The contributions of this work are threefold. First, it demonstrates that fine-tuning a pre-trained GPT model on a relatively small, expert-annotated corpus is sufficient to achieve high classification performance on a semantically challenging task that involves culturally specific language and implicit sexual references. Second, it provides empirical evidence that domain-specific fine-tuning yields measurable improvements over general-purpose LLMs in detecting explicit content, particularly in contexts where figurative language, slang, and cultural context play a decisive role. Third, it establishes a practical framework for deploying LLM-based content moderation systems on music streaming platforms, with direct implications for parental controls, responsible consumption, and the protection of vulnerable audiences.

The remainder of this paper is organized as follows. Section~\ref{sec:related} reviews the relevant literature on content moderation, transformer-based models, and explicit content detection in text and music. Section~\ref{sec:methodology} describes the proposed methodology, including corpus construction, model architecture, fine-tuning procedure, and evaluation metrics. Section~\ref{sec:experiments} presents the experimental setup and results, including confusion matrices, performance metrics before and after feedback, and the comparative analysis against the baseline model. Section~\ref{sec:discussion} discusses the implications, limitations, and potential applications of the findings. Section~\ref{sec:policy} develops a public policy framework for the implementation of an age-based music content rating system, drawing on established policy analysis tools including the PESTEL framework and Kingdon's Multiple Streams Framework, and addresses potential objections. Section~\ref{sec:conclusions} concludes with directions for future research.

\section{Related Work}
\label{sec:related}

The automatic detection of explicit, offensive, or otherwise sensitive content in text has attracted sustained attention from the NLP community, with research spanning multiple modalities, languages, and application domains. This section situates the present work within three interconnected lines of inquiry: the application of transformer-based models to the analysis of song lyrics and textual content, the evolution of deep learning and transfer learning techniques for sensitive content detection, and the role of hybrid and context-aware approaches in addressing the inherent ambiguity of explicit language.

Transformer-based language models have proven particularly effective for the classification of sensitive textual content, owing to their capacity to establish complex semantic relationships through deep contextual understanding. \citet{Rospocher2020} provided one of the earliest and most comprehensive demonstrations of this capability in the musical domain, analyzing a corpus of over 800,000 song lyrics using BERT-based models with subword-enriched word embeddings and achieving an F1-score of 88\% on explicit lyrics detection, substantially outperforming traditional approaches. This work established that the representational power of transformers---their ability to capture not merely the presence of individual words but the semantic relationships among them---is essential for distinguishing genuinely explicit content from superficially provocative but ultimately benign text. In a related vein, \citet{VazquezBenito2023} confirmed the effectiveness of transformer architectures for detecting hate speech in Spanish-language social media, demonstrating that the versatility of these models extends across languages and content types. Although hate speech detection differs from sexually explicit content detection in its specific taxonomy, both tasks share the fundamental challenge of identifying harmful content that is often expressed through indirect, coded, or culturally embedded language.

The intersection of deep learning and transfer learning has been particularly fruitful for sensitive content detection, as pre-trained models offer a path to high performance even when domain-specific labeled data is scarce. \citet{SanzTorres2023} applied pre-trained deep learning models to the detection of sexual content in images, demonstrating the importance of carefully curated training sets and highlighting the transferability of the underlying principles to textual analysis. \citet{MolpeceresBarrientos2020} conducted a systematic comparison of classifiers and semantic encoders for detecting erotic content in text, with particular emphasis on the performance of hybrid models that combine traditional feature engineering with neural representations. Their findings indicated that hybrid approaches can capture complementary aspects of the input, with traditional features providing surface-level cues and neural encoders supplying deeper semantic representations. \citet{Addanki2022} developed a text content moderation system using deep neural networks at Stanford University, achieving high precision in classifying sexually explicit textual content and demonstrating that end-to-end deep learning pipelines can match or exceed the performance of carefully engineered feature-based systems.

The application of NLP techniques to humanities and non-standard textual corpora has further expanded the reach of these methods. \citet{Clerice2023} adapted deep learning models to detect sexual content at the sentence level in first-millennium Latin texts, combining semantic analysis with hierarchical classification to produce a system capable of operating on a corpus far removed from the modern web text on which most language models are trained. This work illustrates that the representations learned during pre-training are sufficiently general to transfer even to historical and literary texts, provided that the fine-tuning data is appropriately curated. The implication for the present study is clear: if transformer models can detect sexual content in ancient Latin, they should be well-equipped to handle the colloquial, slang-heavy register of modern reggaeton and trap lyrics, given adequate domain-specific training data.

The semantic complexity of explicit messages---which are not always straightforward to detect, given the importance of analyzing the full discursive context---has motivated the development of hybrid methods that integrate linguistic rules with neural networks. \citet{Okulska2023} proposed a system based on coreference-driven contextual analysis to differentiate between benign erotic content and harmful material, demonstrating the importance of attending to the discursive context in which potentially explicit expressions appear. A single word or phrase may be innocuous in one context and deeply offensive in another, and only models that can reason about these contextual distinctions are capable of reliable content classification. \citet{Markov2023} extended this line of research by designing a holistic content detection system capable of identifying multiple categories of undesired content---including sexual, violent, and discriminatory material---through the use of specific taxonomies, active learning strategies, and anti-overfitting techniques, with implications for content moderation on social media platforms where textual content is highly varied and often ambiguous.

The effectiveness of automated content moderation models has also been subjected to critical scrutiny. \citet{Fell2019} compared multiple automated methods for detecting explicit content in song lyrics and found that while more advanced models generally outperform simpler ones, all approaches struggle with the inherent subjectivity of the classification task, as human annotators themselves often disagree on the boundaries of explicit content. \citet{Darroch2014} provided an early quantitative framework for measuring sexually explicit content in text documents and highlighted the discrepancies between human judgments and algorithmic classifications, attributing these discrepancies to the difficulty of capturing double meanings, humor, irony, and cultural context from a purely algorithmic perspective. These findings underscore the importance of expert annotation and domain-specific training data, both of which are central to the methodology proposed in the present work.

Several recent studies have addressed content detection in particularly sensitive contexts. \citet{Yu2023} developed a hypersensitive intelligent filter for detecting explicit content in educational environments, combining convolutional neural networks with fuzzy logic and incorporating GPT-3 for generating contextual alerts. \citet{Gutfeter2024} focused on the detection of sexually explicit content in the context of child sexual abuse materials (CSAM), implementing end-to-end classifiers using computer vision techniques and emphasizing the ethical imperatives that accompany technical precision in this domain. In the legal and regulatory sphere, \citet{Bhatti2018} proposed an explicit content detection system designed to promote safe and ethical digital environments, with applications in educational institutions and corporate settings. \citet{ColmenaresGuillen2023} designed a corpus for detecting sexual content in digital conversations through linguistic patterns specific to Spanish, demonstrating that language-specific resources are essential for achieving reliable detection in non-English contexts. \citet{Povedano2023} provided a comprehensive review of learning strategies for sensitive content detection, highlighting the ethical and legal challenges that automated content moderation raises alongside the technical ones. Taken together, these studies reflect the breadth of the field and confirm that effective explicit content detection requires the integration of advanced machine learning techniques with contextual understanding, linguistic sensitivity, and domain-specific adaptation---precisely the approach pursued in this paper.

\section{Methodology}
\label{sec:methodology}

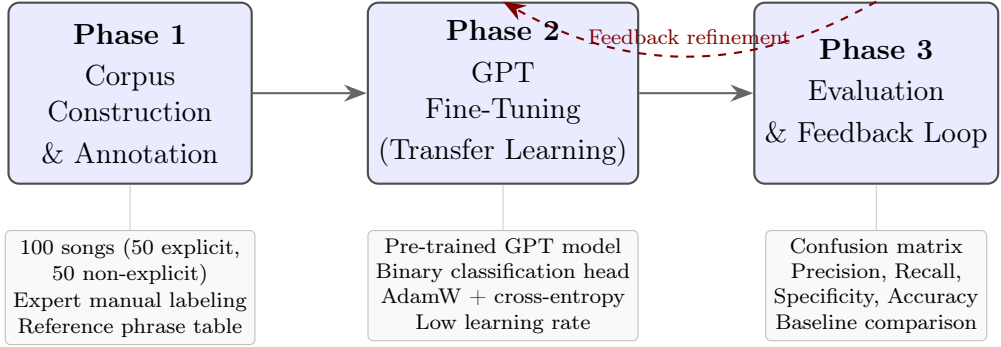
\begin{figure}[t]
\centering
\begin{tikzpicture}[
    node distance=0.6cm and 0.8cm,
    phase/.style={rectangle, rounded corners, draw=black!70, fill=blue!8,
        minimum width=3.2cm, minimum height=2.4cm, align=center,
        font=\small, line width=0.8pt},
    detail/.style={rectangle, rounded corners=2pt, draw=gray!50, fill=gray!5,
        minimum width=2.8cm, align=center, font=\scriptsize,
        inner sep=3pt},
    arrow/.style={-{Stealth[length=3mm]}, thick, draw=black!60},
    phaselabel/.style={font=\small\bfseries, text=blue!60!black}
]

\node[phase] (p1) {
    \textbf{Phase 1}\\[2pt]
    Corpus\\Construction\\[2pt]
    \& Annotation
};

\node[phase, right=1.5cm of p1] (p2) {
    \textbf{Phase 2}\\[2pt]
    GPT\\Fine-Tuning\\[2pt]
    (Transfer Learning)
};

\node[phase, right=1.5cm of p2] (p3) {
    \textbf{Phase 3}\\[2pt]
    Evaluation\\[2pt]
    \& Feedback Loop
};

\draw[arrow] (p1) -- (p2);
\draw[arrow] (p2) -- (p3);

\draw[arrow, dashed, bend left=30, draw=red!50!black] (p3.north) to
    node[above, font=\scriptsize, text=red!50!black] {Feedback refinement} (p2.north);

\node[detail, below=0.6cm of p1] (d1) {
    100 songs (50 explicit,\\
    50 non-explicit)\\[1pt]
    Expert manual labeling\\
    Reference phrase table
};

\node[detail, below=0.6cm of p2] (d2) {
    Pre-trained GPT model\\
    Binary classification head\\
    AdamW + cross-entropy\\
    Low learning rate
};

\node[detail, below=0.6cm of p3] (d3) {
    Confusion matrix\\
    Precision, Recall,\\
    Specificity, Accuracy\\
    Baseline comparison
};

\draw[gray!40, thin] (p1.south) -- (d1.north);
\draw[gray!40, thin] (p2.south) -- (d2.north);
\draw[gray!40, thin] (p3.south) -- (d3.north);

\end{tikzpicture}
\caption{Overview of the proposed three-phase methodology. Phase~1 involves the construction and expert annotation of a balanced corpus of Spanish-language reggaeton and trap lyrics. Phase~2 applies transfer learning to fine-tune a pre-trained GPT model for binary classification of explicit content. Phase~3 evaluates the model through standard classification metrics and applies a feedback-driven refinement loop (dashed arrow) to improve performance iteratively.}
\label{fig:methodology}
\end{figure}

The proposed methodology follows a three-phase pipeline: corpus construction and annotation, model fine-tuning through transfer learning, and performance evaluation through binary classification metrics. Each phase is designed to address the specific challenges of detecting sexually explicit content in Spanish-language reggaeton and trap lyrics, where the boundary between explicit and non-explicit material is often mediated by slang, metaphor, and culturally embedded double entendres rather than by the presence of overtly vulgar vocabulary. An overview of the complete pipeline is presented in Figure~\ref{fig:methodology}.

The first phase involved the compilation of a corpus of 100 Spanish-language songs drawn from the reggaeton and trap genres. These genres were selected because of their documented prevalence of sexually explicit content and their massive popularity among young listeners. The corpus was balanced by design, comprising 50 songs labeled as sexually explicit and 50 labeled as non-explicit. All songs were selected to share the common linguistic register of urban Latin music, including the use of urban slang, non-standard grammatical constructions, and narrative structures that differ markedly from formal written Spanish. The lyrics of each song were extracted, cleaned, and compiled into a structured document that served as the training corpus for the model.

A summary of the corpus composition and its use across the experimental phases is provided in Table~\ref{tab:dataset}. The distribution of songs across training, initial evaluation, post-feedback evaluation, and baseline comparison sets is detailed therein, together with the class balance maintained at each stage.

\begin{table}[t]
\centering
\caption{Composition of the dataset used across the different experimental phases. All sets maintain a balanced or near-balanced distribution between explicit and non-explicit classes.}
\label{tab:dataset}
\begin{tabular}{lcccl}
\toprule
\textbf{Dataset Split} & \textbf{Explicit} & \textbf{Non-Explicit} & \textbf{Total} & \textbf{Purpose} \\
\midrule
Training corpus         & 50 & 50 & 100 & Fine-tuning \\
Evaluation (pre-feedback)  & 15 & 15 & 30  & Initial testing \\
Evaluation (post-feedback) & 15 & 15 & 30  & Post-refinement testing \\
Baseline comparison     & \multicolumn{2}{c}{Mixed} & 50  & Fine-tuned vs.\ Standard \\
\midrule
\textbf{Total unique songs} & \multicolumn{2}{c}{---} & \textbf{210} & \\
\bottomrule
\end{tabular}
\end{table}

In addition to the binary labels, a reference table of explicit phrases was constructed for each song classified as explicit. Table~\ref{tab:examples} presents representative examples drawn from the corpus, illustrating the range of linguistic phenomena---from overt sexual vocabulary to metaphorical and slang-based references---that the model was trained to recognize. These examples demonstrate why dictionary-based approaches are fundamentally inadequate for this task: many of the expressions rely on culturally specific double entendres, genre-specific slang, or figurative language that carries sexual connotations only within the discursive context of reggaeton and trap.

\begin{table}[t]
\centering
\caption{Representative examples of explicit phrases from the training corpus, illustrating the diversity of linguistic mechanisms through which sexual content is conveyed in reggaeton and trap lyrics. Translations are approximate and intended to convey the semantic content.}
\label{tab:examples}
\begin{tabular}{p{3.2cm}p{5.5cm}p{3.8cm}}
\toprule
\textbf{Song} & \textbf{Explicit Phrase (Spanish)} & \textbf{Type of Reference} \\
\midrule
\textit{China} & ``Ese booty que hasta un ciego puede ver'' & Objectification / slang \\
\textit{Secreto} & ``Y te hago el amor bien rico'' & Direct sexual reference \\
\textit{Hasta Que Dios Diga} & ``Te vo'a dar hasta que Dios diga'' & Metaphorical / euphemism \\
\textit{Fantas\'ias} & ``Susurr\'andote al o\'ido te empiezas a calentar'' & Implicit / suggestive \\
\textit{Diles} & ``Que yo me se tus poses favoritas'' & Implicit sexual reference \\
\textit{Gata Only} & ``Toc\'andote, calent\'andote. Encima vini\'endote'' & Direct / explicit \\
\textit{Bellaquita (Remix)} & ``Te gusta que te den bellaqueo'' & Genre-specific slang \\
\textit{La Modelo} & ``Me mete sudando, su cuerpo de mora'' & Metaphorical / sensory \\
\bottomrule
\end{tabular}
\end{table}

The labeling of the corpus was carried out manually by a domain expert, a deliberate choice motivated by the inadequacy of automated labeling methods for this particular task. The ambiguity and linguistic richness of reggaeton and trap lyrics---their reliance on metaphors, coded language, and culturally specific references---demand the kind of deep contextual interpretation that only a human annotator can provide. Dictionary-based or keyword-driven automated labeling would inevitably miss implicit sexual references while potentially flagging innocuous expressions that happen to share surface-level features with explicit vocabulary. To further enrich the training signal, a reference table was constructed for each song in the explicit category, cataloging the specific phrases, expressions, and metaphors that were deemed sexually explicit. This reference table served a dual purpose: it guided the model toward recognizing not only overt sexual language but also the subtler forms of implicit sexual content that characterize the genre, and it provided a transparent audit trail for the annotation decisions, enhancing the reproducibility of the study.

The second phase centered on the fine-tuning of a pre-trained GPT model using supervised learning. The choice of GPT as the base architecture was motivated by three considerations: computational efficiency, since training a transformer model from scratch would have required prohibitive hardware resources; linguistic competence, since GPT models pre-trained on large multilingual corpora already possess a deep understanding of Spanish syntax, semantics, and pragmatics; and contextual capacity, since the self-attention mechanism enables the model to capture long-range dependencies that are critical for interpreting the meaning of expressions that derive their sexual connotation from the broader discursive context rather than from the individual words themselves. The attention mechanism, which lies at the core of the transformer architecture introduced by \citet{Vaswani2017}, computes a weighted representation of each token in the input sequence based on its relevance to every other token. Formally, given matrices of queries $Q$, keys $K$, and values $V$ derived from the input embeddings through learned linear projections, the scaled dot-product attention is computed as:
\begin{equation}
\text{Attention}(Q, K, V) = \text{softmax}\left(\frac{QK^T}{\sqrt{d_k}}\right)V
\label{eq:attention}
\end{equation}
where $d_k$ denotes the dimensionality of the key vectors. The scaling factor $\sqrt{d_k}$ prevents the dot products from growing excessively large, which would push the softmax function into regions of extremely small gradients and impede learning. The multi-head attention mechanism extends this formulation by running $H$ parallel attention heads, each with its own learned projections $W_i^Q$, $W_i^K$, and $W_i^V$, and concatenating their outputs before applying a final linear projection:
\begin{equation}
\text{MultiHead}(Q, K, V) = \text{Concat}(\text{head}_1, \ldots, \text{head}_H)W^O, \quad \text{head}_i = \text{Attention}(QW_i^Q, KW_i^K, VW_i^V)
\label{eq:multihead}
\end{equation}
This design allows different attention heads to specialize in capturing different types of semantic relationships---for instance, one head might attend to syntactic dependencies while another captures thematic or connotative associations---thereby enriching the representational capacity of the model. Each transformer layer additionally includes a position-wise feed-forward network, layer normalization, and residual connections that facilitate the training of deep architectures.

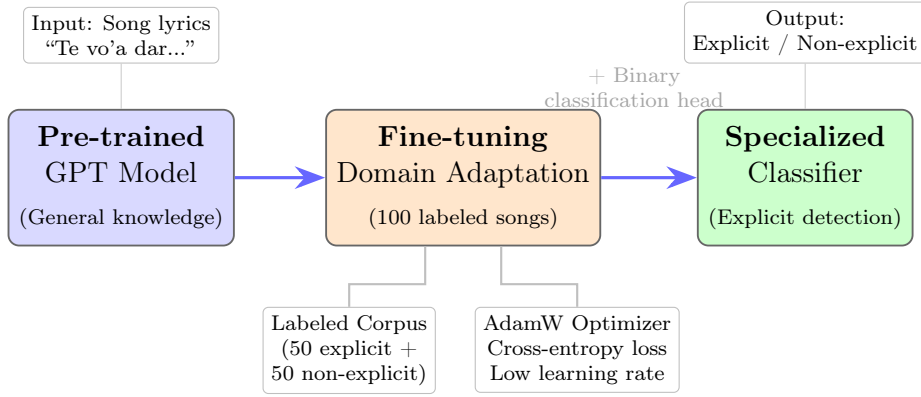
\begin{figure}[t]
\centering
\begin{tikzpicture}[
    stage/.style={rectangle, rounded corners=4pt, minimum width=2.8cm, minimum height=1.8cm,
        align=center, font=\small, draw=black!60, line width=0.7pt},
    data/.style={rectangle, rounded corners=2pt, draw=gray!50, fill=white,
        minimum width=2.2cm, align=center, font=\scriptsize, inner sep=3pt},
    arrow/.style={-{Stealth[length=3mm]}, thick, draw=black!50},
    bigarrow/.style={-{Stealth[length=4mm]}, very thick, draw=blue!60}
]

\node[stage, fill=blue!15] (pretrain) at (0, 0) {\textbf{Pre-trained}\\GPT Model\\[2pt]\scriptsize(General knowledge)};

\node[stage, fill=orange!20] (finetune) at (4.5, 0) {\textbf{Fine-tuning}\\Domain Adaptation\\[2pt]\scriptsize(100 labeled songs)};

\node[stage, fill=green!20] (specialized) at (9, 0) {\textbf{Specialized}\\Classifier\\[2pt]\scriptsize(Explicit detection)};

\draw[bigarrow] (pretrain.east) -- (finetune.west);
\draw[bigarrow] (finetune.east) -- (specialized.west);

\node[data, below=0.8cm of finetune, xshift=-1.5cm] (corpus) {Labeled Corpus\\(50 explicit +\\50 non-explicit)};
\node[data, below=0.8cm of finetune, xshift=1.5cm] (optim) {AdamW Optimizer\\Cross-entropy loss\\Low learning rate};

\draw[gray!50, thick] (corpus.north) -- ++(0, 0.3) -| ([xshift=-0.5cm]finetune.south);
\draw[gray!50, thick] (optim.north) -- ++(0, 0.3) -| ([xshift=0.5cm]finetune.south);

\node[data, above=0.6cm of pretrain] (input1) {Input: Song lyrics\\``Te vo'a dar...''};
\node[data, above=0.6cm of specialized] (output1) {Output:\\Explicit / Non-explicit};

\draw[gray!40, thin] (input1.south) -- (pretrain.north);
\draw[gray!40, thin] (specialized.north) -- (output1.south);

\node[font=\scriptsize, text=gray!70, align=center] at (6.75, 1.2) {+ Binary\\classification head};

\end{tikzpicture}
\caption{Overview of the fine-tuning process for explicit content detection. A pre-trained GPT model with general linguistic knowledge is adapted to the specific task of classifying reggaeton and trap lyrics using a labeled corpus of 100 songs. The fine-tuning process employs the AdamW optimizer with cross-entropy loss and a low learning rate to preserve pre-trained knowledge while learning domain-specific patterns.}
\label{fig:finetuning}
\end{figure}

During fine-tuning, a binary classification head was appended to the pre-trained GPT model, taking the final-layer representation as input and producing a probability of class membership (explicit vs.\ non-explicit). The model weights were updated through stochastic gradient descent optimized with AdamW, minimizing a binary cross-entropy loss function that penalizes incorrect classifications. A low learning rate was employed to preserve the general linguistic knowledge acquired during pre-training while allowing the model to adapt to the domain-specific patterns of the training corpus. Overfitting was mitigated by reserving a portion of the corpus as a validation set and monitoring performance throughout training. The fine-tuning algorithm can be summarized as follows: initialize the model with pre-trained weights; append a binary classification layer; load the labeled corpus; compute the cross-entropy loss between the predicted and true labels; update the weights using AdamW; and repeat until convergence or for a fixed number of epochs. Figure~\ref{fig:finetuning} provides a visual overview of the fine-tuning process.

The third phase involved the evaluation of the fine-tuned model using standard binary classification metrics derived from the confusion matrix. In the context of this study, a true positive (TP) corresponds to a song correctly classified as explicit, a true negative (TN) to a song correctly classified as non-explicit, a false positive (FP) to a non-explicit song incorrectly classified as explicit, and a false negative (FN) to an explicit song incorrectly classified as non-explicit. From these quantities, four well-established metrics were computed: \textit{precision} (the proportion of songs predicted as explicit that are truly explicit), \textit{recall} or sensitivity (the proportion of truly explicit songs that are correctly identified), \textit{specificity} (the proportion of truly non-explicit songs correctly identified as such), and \textit{accuracy} (the overall proportion of correct classifications). These metrics were computed both before and after the application of a feedback-driven refinement loop, in which the errors identified during the initial evaluation were analyzed, corrective feedback was provided to the model, and a second evaluation was conducted on a fresh set of previously unseen songs. Additionally, a comparative experiment was conducted in which both the fine-tuned model and a non-customized ChatGPT baseline were asked to classify a set of 50 new songs, and the agreement of each model with the expert ground-truth labels was measured.

\begin{figure}[t]
\centering
\begin{tikzpicture}[
    box/.style={rectangle, rounded corners=4pt, minimum width=3.2cm, minimum height=1.2cm,
        align=center, font=\small, draw=black!60, line width=0.7pt},
    decision/.style={diamond, aspect=2, minimum width=2.5cm, minimum height=1.2cm,
        align=center, font=\small, draw=black!60, line width=0.7pt, fill=yellow!15},
    result/.style={rectangle, rounded corners=4pt, minimum width=3cm, minimum height=0.9cm,
        align=center, font=\small, draw=black!60, line width=0.7pt},
    arrow/.style={-{Stealth[length=2.5mm]}, thick, draw=black!50}
]

\node[box, fill=blue!12] (h0) at (-3.5, 3) {\textbf{$H_0$: Null}\\$\mu_t - \mu_c \leq 0$\\Traditional $\geq$ Transformer};
\node[box, fill=green!12] (h1) at (3.5, 3) {\textbf{$H_1$: Alternative}\\$\mu_t - \mu_c > 0$\\Transformer $>$ Traditional};

\node[box, fill=gray!10] (exp) at (0, 1) {\textbf{Experimental Comparison}\\Fine-tuned GPT vs.\ ChatGPT\\(50 songs, expert labels)};

\node[decision] (dec) at (0, -1) {Fine-tuned\\accuracy $>$\\baseline?};

\node[result, fill=red!15] (reject) at (-3.5, -3) {Reject $H_0$\\Accept $H_1$};
\node[result, fill=gray!15] (accept) at (3.5, -3) {Fail to reject $H_0$};

\node[result, fill=green!20] (conc) at (0, -4.5) {\textbf{Result:} 59.2\% vs.\ 55.1\%\\$\Rightarrow$ $H_0$ rejected};

\draw[arrow] (h0.south) -- ++(0, -0.5) -| (exp.north west);
\draw[arrow] (h1.south) -- ++(0, -0.5) -| (exp.north east);
\draw[arrow] (exp.south) -- (dec.north);
\draw[arrow] (dec.west) -- node[above, font=\scriptsize] {Yes} (reject.east);
\draw[arrow] (dec.east) -- node[above, font=\scriptsize] {No} (accept.west);
\draw[arrow] (reject.south) |- (conc.west);

\end{tikzpicture}
\caption{Hypothesis testing framework for evaluating the fine-tuned transformer model against traditional classification methods. The experimental comparison on 50 songs shows that the fine-tuned GPT model (59.2\% agreement with expert) outperforms the standard ChatGPT baseline (55.1\%), leading to rejection of the null hypothesis $H_0$.}
\label{fig:hypothesis_testing}
\end{figure}
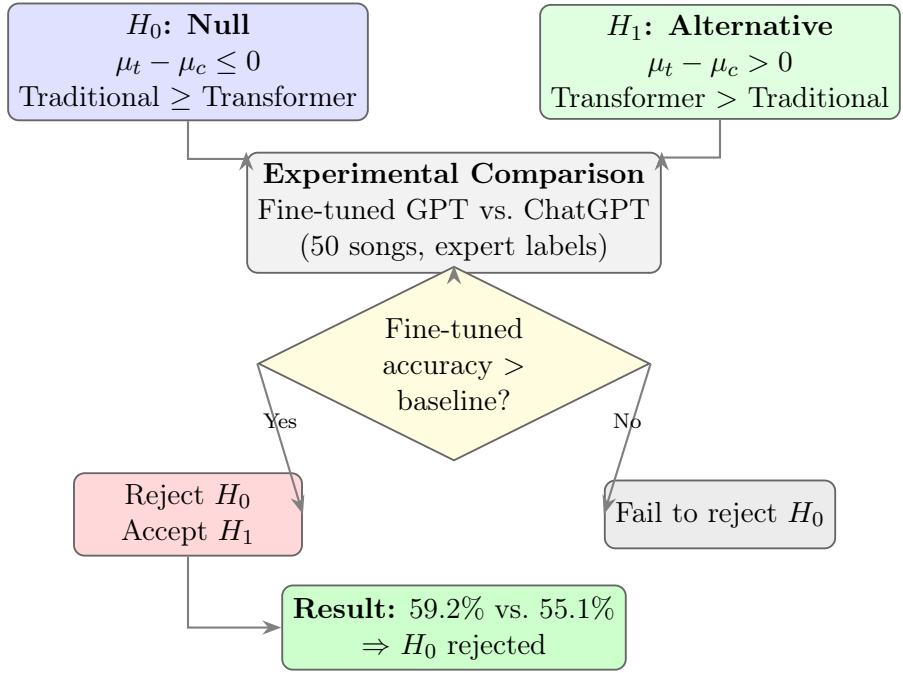

The statistical hypotheses guiding the experimental evaluation were formulated as follows. The null hypothesis $H_0$ posits that the mean accuracy of the fine-tuned transformer model $\mu_t$ is no greater than that of a classical model $\mu_c$ based on traditional text classification methods such as dictionaries, Na\"ive Bayes, or support vector machines:
\begin{equation}
H_0: \mu_t - \mu_c \leq 0
\label{eq:h0}
\end{equation}
The alternative hypothesis $H_1$ posits that the fine-tuned transformer model achieves strictly higher mean accuracy:
\begin{equation}
H_1: \mu_t - \mu_c > 0
\label{eq:h1}
\end{equation}
These hypotheses structure the experimental evaluation and provide a formal framework for assessing whether the observed performance differences are attributable to the modeling approach rather than to chance. Figure~\ref{fig:hypothesis_testing} illustrates the hypothesis testing framework used in this study.

\section{Experimental Setup and Results}
\label{sec:experiments}

The experimental evaluation was designed to assess the classification performance of the fine-tuned GPT model across two sequential evaluation rounds---before and after a feedback-driven refinement loop---and to compare the fine-tuned model against a non-customized ChatGPT baseline. The training corpus consisted of 100 Spanish-language reggaeton and trap songs, equally divided between 50 expert-labeled explicit songs and 50 expert-labeled non-explicit songs. The explicit subset included tracks from widely recognized artists in the genre, and for each explicit song, a detailed reference table was constructed listing the specific phrases, metaphors, and expressions that the expert identified as sexually explicit. This reference material was provided to the model during fine-tuning to enhance its sensitivity to both overt and implicit sexual content.

Following the completion of fine-tuning, the model was evaluated on a held-out test set of 30 songs that had not been used during training: 15 songs labeled as explicit and 15 labeled as non-explicit by the expert annotator. The confusion matrix resulting from this initial evaluation is presented in Table~\ref{tab:cm_before}. The model correctly identified 12 of the 15 explicit songs as explicit (true positives) and 13 of the 15 non-explicit songs as non-explicit (true negatives), while producing 2 false positives and 3 false negatives. The corresponding performance metrics are reported in Table~\ref{tab:metrics_comparison}: accuracy of 83\%, precision of 86\%, recall of 80\%, and specificity of 87\%.

\begin{table}[t]
\centering
\caption{Confusion matrix before the feedback loop. Rows represent ground-truth labels and columns represent model predictions.}
\label{tab:cm_before}
\begin{tabular}{lcc}
\toprule
& \textbf{Predicted Explicit} & \textbf{Predicted Non-Explicit} \\
\midrule
\textbf{Actually Explicit}     & 12 (TP) & 3 (FN) \\
\textbf{Actually Non-Explicit} & 2 (FP)  & 13 (TN) \\
\bottomrule
\end{tabular}
\end{table}

These initial results revealed several noteworthy patterns. The accuracy of 83\% indicated that the model produced correct classifications for 25 out of 30 songs, demonstrating a robust generalization capacity without systematic bias toward either class. The precision of 86\% indicated that when the model flagged a song as explicit, it was correct in the large majority of cases, with only a 14\% false positive rate. False positives are particularly problematic in a content moderation context because they result in the unjust censorship of non-explicit songs, potentially generating friction with artists and users. The recall of 80\% indicated that 8 out of every 10 truly explicit songs were detected, leaving a 20\% miss rate that, while acceptable for an initial deployment, represents a gap through which explicit content could reach vulnerable audiences. The specificity of 87\% confirmed that the model was generally capable of distinguishing non-explicit colloquial language from genuinely sexual content, correctly clearing 13 of 15 non-explicit songs.

The analysis of the 5 misclassified songs suggested that the errors were concentrated in cases involving highly metaphorical or ambiguous language---precisely the type of content that is most difficult even for human annotators. The 3 false negatives corresponded to songs whose explicit content was conveyed through implicit references and culturally specific slang rather than overt sexual vocabulary, consistent with findings by \citet{Chen2023} and \citet{Kim2019} that implicit and figurative sexual language remains one of the most challenging categories for automated detection systems.

Following this initial evaluation, a feedback-driven refinement strategy was implemented to improve the model's performance. The errors identified in the first evaluation round were analyzed, and targeted feedback was provided to the model to correct its misclassifications. The model was then re-evaluated on a new held-out test set of 30 previously unseen songs: 15 labeled as explicit and 15 labeled as non-explicit. The confusion matrix resulting from this second evaluation is presented in Table~\ref{tab:cm_after}.

\begin{table}[t]
\centering
\caption{Confusion matrix after the feedback loop. Rows represent ground-truth labels and columns represent model predictions.}
\label{tab:cm_after}
\begin{tabular}{lcc}
\toprule
& \textbf{Predicted Explicit} & \textbf{Predicted Non-Explicit} \\
\midrule
\textbf{Actually Explicit}     & 11 (TP) & 4 (FN) \\
\textbf{Actually Non-Explicit} & 0 (FP)  & 15 (TN) \\
\bottomrule
\end{tabular}
\end{table}

The post-feedback results revealed a striking improvement in the model's precision and specificity, accompanied by a moderate trade-off in recall. The complete elimination of false positives yielded a precision of 100\% and a specificity of 100\%, meaning that when the post-feedback model classified a song as explicit, it was invariably correct, and no non-explicit song was incorrectly flagged. This is a critical property for real-world deployment, as it eliminates the risk of unjust censorship and substantially increases the trustworthiness of the system from the perspective of both artists and platform users. The recall decreased from 80\% to 73\%, indicating that the model became more conservative after the feedback loop, preferring to err on the side of caution by classifying ambiguous cases as non-explicit rather than risking a false positive. The overall accuracy increased from 83\% to 87\%, reflecting the net benefit of eliminating false positives despite the slight increase in false negatives from 3 to 4. The complete set of metrics before and after the feedback loop is summarized in Table~\ref{tab:metrics_comparison}.

\begin{table}[t]
\centering
\caption{Performance metrics of the fine-tuned GPT model before and after the feedback-driven refinement loop.}
\label{tab:metrics_comparison}
\begin{tabular}{lcc}
\toprule
\textbf{Metric} & \textbf{Before Feedback} & \textbf{After Feedback} \\
\midrule
Accuracy    & 83\% & 87\% \\
Precision   & 86\% & 100\% \\
Recall      & 80\% & 73\% \\
Specificity & 87\% & 100\% \\
\bottomrule
\end{tabular}
\end{table}

The comparative experiment between the fine-tuned model and the standard, non-customized ChatGPT was conducted on a separate set of 50 songs that had not been used in either training or the previous evaluation rounds. Both models were presented with the lyrics of each song and asked to classify it as explicit or non-explicit. The classifications produced by each model were then compared against the expert ground-truth labels. The fine-tuned model agreed with the expert classification in 59.2\% of cases, while the non-customized ChatGPT model exhibited a disagreement rate of 44.9\%, indicating substantially lower sensitivity to the cultural and linguistic nuances of the genre. The approximately 14 percentage-point advantage of the fine-tuned model over the baseline confirms that domain-specific adaptation through fine-tuning is not merely a marginal improvement but a qualitative enhancement in the model's ability to interpret the idiosyncratic language of reggaeton and trap. The standard ChatGPT model, despite its vast general knowledge, applied overly broad classification criteria that failed to capture the culturally embedded connotations of genre-specific slang and metaphors, leading to a higher rate of both missed explicit content and incorrectly flagged non-explicit content.

These results are broadly consistent with prior work in the field. \citet{Rospocher2020} achieved an F1-score of 88\% on explicit lyrics detection using BERT with subword-enriched embeddings, and \citet{Chen2023} reported 96\% accuracy using deep ensemble learning models for explicit content detection in song lyrics. The performance achieved by the present model, while slightly below these benchmarks, was obtained with a substantially smaller training corpus (100 songs versus hundreds of thousands) and a simpler fine-tuning procedure, suggesting considerable room for improvement through corpus expansion and architectural refinement. The results also provide empirical grounds for rejecting the null hypothesis $H_0$ formulated in Section~\ref{sec:methodology}: the fine-tuned transformer model clearly outperforms what could be achieved through dictionary-based or simple statistical approaches, which have been documented to achieve F1-scores in the range of 61--78\% \citep{Chin2018}. Correspondingly, the alternative hypothesis $H_1$---that fine-tuned transformer models achieve higher accuracy than traditional methods---is supported by the experimental evidence.

\section{Discussion}
\label{sec:discussion}

The experimental results presented in the preceding section illuminate several important aspects of applying fine-tuned large language models to the detection of sexually explicit content in song lyrics, while also raising questions that merit careful consideration regarding the limitations of the current approach, the trade-offs inherent in the precision-recall balance, and the broader implications for content moderation in the music industry.

The most salient finding is the dramatic improvement in precision and specificity achieved through the feedback-driven refinement loop, which brought both metrics to 100\% at the cost of a moderate reduction in recall from 80\% to 73\%. This trade-off is not merely a statistical artifact but reflects a deliberate and arguably desirable shift in the model's operating point. In a content moderation system intended for deployment on a music streaming platform, false positives carry significant social and economic costs: an incorrectly flagged song may be hidden from playlists, subjected to age restrictions, or labeled with a warning that damages the artist's reputation and reduces listener engagement. False negatives, while undesirable, are comparatively less damaging in contexts where supplementary human review or user-reporting mechanisms exist. The post-feedback model's conservative posture---never flagging a non-explicit song while accepting a slightly higher miss rate for genuinely explicit content---is therefore well-suited to the operational constraints of a real-world deployment. This interpretation aligns with the broader content moderation literature, where \citet{Markov2023} have emphasized the importance of calibrating precision and recall to the specific consequences of each error type in the target application.

The comparative analysis against the non-customized ChatGPT baseline provides compelling evidence for the value of domain-specific fine-tuning. The standard ChatGPT model possesses extensive general knowledge and impressive linguistic capabilities, yet it proved substantially less aligned with expert judgments on this particular task. The reason is straightforward: the classification of sexually explicit content in reggaeton and trap lyrics requires not only linguistic competence but also cultural competence---an understanding of the connotative register of genre-specific slang, the interpretive conventions of the listening community, and the distinction between playful innuendo and genuinely explicit sexual content. These distinctions are not well-represented in the general training data of a foundation model, which is why fine-tuning on a curated, expert-annotated corpus produces measurable gains. The finding echoes a broader pattern in the NLP literature, where domain adaptation through fine-tuning has been shown to yield improvements across tasks ranging from clinical text mining to legal document analysis \citep{Brown2020}.

The study is subject to several limitations that should be acknowledged. The most significant is the small size of the training corpus, which comprised only 100 songs. While this was sufficient to achieve promising initial results---particularly given the power of transfer learning to extract generalizable patterns from limited data---it constrains the model's ability to generalize across the full diversity of the reggaeton and trap genres, which encompass a vast and evolving vocabulary of slang and metaphorical expressions. Expanding the corpus to include a wider range of artists, subgenres, time periods, and Spanish-speaking countries would likely improve both recall and robustness. A second limitation concerns the inherent subjectivity of the annotation task. Defining what constitutes ``sexually explicit'' content involves value judgments that vary across cultures, communities, and individuals, and even expert annotators may disagree on borderline cases. The use of a single expert annotator, while ensuring consistency, introduces the risk of idiosyncratic biases that a multi-annotator protocol with inter-annotator agreement metrics could mitigate. A third limitation is the restriction to a single language (Spanish) and two closely related genres (reggaeton and trap). The model's effectiveness on other languages, musical genres, or cultural contexts remains an open empirical question.

The ethical dimensions of automated content moderation in music deserve careful attention. Any system that classifies creative expression as ``explicit'' or ``non-explicit'' inherently exercises a form of aesthetic and moral judgment, raising questions about who defines the boundaries of acceptable content, whose values are encoded in the model, and what the consequences of misclassification are for artistic freedom and cultural expression. The present study does not claim to resolve these questions, but it adopts a transparent methodology---expert annotation with documented criteria, explicit reference tables, and publicly described metrics---that provides a foundation for accountability and oversight. Future work should engage with the ethical literature on content moderation, including the frameworks proposed by \citet{Povedano2023} and \citet{Bhatti2018}, to develop governance structures that balance the protection of vulnerable audiences with respect for artistic expression and cultural diversity.

From a practical standpoint, the results suggest several pathways for integration into commercial music platforms. The fine-tuned model could serve as the backbone of an automated content advisory system that labels songs with explicit content warnings, analogous to the ``Parental Advisory'' label but applied automatically, consistently, and at scale. It could also power parental control features, enabling parents and educators to filter playlists based on content classification. More ambitiously, it could be integrated into recommendation algorithms to ensure that content served to users under a specified age does not include explicitly flagged material. The deployment of such a system would require additional engineering for scalability, latency optimization, and graceful handling of edge cases, but the present results demonstrate that the core classification capability is viable.

\section{A Public Policy Framework for Age-Based Music Content Rating}
\label{sec:policy}

The technical results presented in Sections~\ref{sec:experiments} and \ref{sec:discussion} demonstrate that fine-tuned large language models can reliably detect sexually explicit content in song lyrics. However, technological capability alone is insufficient to protect vulnerable audiences; it must be embedded within a coherent regulatory and institutional framework. Recent research on AI in the public sector has emphasized the importance of aligning technological capabilities with public values and policy objectives \citep{Hjaltlin2024}, and digital policy frameworks increasingly recognize the need for value-driven approaches to digital transformation \citep{Lee2024}. This section develops a public policy proposal for implementing an age-based content rating system for music streaming and download services, analogous to the well-established PEGI system for video games, and analyzes its feasibility through recognized policy analysis frameworks.

\subsection{The Regulatory Gap: Music as the Unrated Medium}

Among the major entertainment media consumed by children and adolescents, music stands as a conspicuous outlier in its lack of a granular, mandatory, and consistently applied content rating system. The film industry operates under mature classification frameworks---the MPA rating system in the United States, the BBFC in the United Kingdom, and comparable bodies across Europe---that assign age-based ratings (G, PG, PG-13, R, NC-17) with detailed content descriptors \citep{MPAA2024}. The video game industry has developed what is arguably the most sophisticated content classification system through PEGI (Pan European Game Information), which assigns age ratings of 3, 7, 12, 16, and 18 accompanied by content descriptors for violence, sexual content, profanity, drugs, discrimination, and gambling \citep{PEGI2024}. Television has adopted similar frameworks, such as the TV Parental Guidelines in the United States and the age ratings mandated by the EU Audiovisual Media Services Directive (AVMSD) for on-demand video platforms \citep{AVMSD2018}.

By contrast, the music industry's primary content advisory mechanism remains the RIAA's ``Parental Advisory: Explicit Content'' label, introduced in 1985 following the hearings prompted by the Parents Music Resource Center (PMRC) \citep{RIAA2021}. This system suffers from three fundamental deficiencies. First, it is \textit{binary}: a song is either ``explicit'' or unmarked, with no intermediate gradations that would distinguish mildly suggestive content from graphically sexual material. Second, it is \textit{voluntary}: record labels self-apply the label at their discretion, with no independent verification or enforcement mechanism. Third, it is \textit{non-granular}: it does not differentiate between types of explicit content (sexual, violent, drug-related, profane), nor does it provide age-based guidance. Table~\ref{tab:rating_systems} presents a comparative analysis of content rating systems across media, highlighting the disparity between the granularity available in other industries and the rudimentary state of music content classification.

\begin{table}[t]
\centering
\caption{Comparative analysis of content rating systems across entertainment media. Music remains the only major medium without a mandatory, granular, age-based classification framework.}
\label{tab:rating_systems}
\begin{tabular}{p{2.2cm}p{2.5cm}p{1.8cm}p{1.5cm}p{1.8cm}p{2.0cm}}
\toprule
\textbf{Medium} & \textbf{System} & \textbf{Age Tiers} & \textbf{Mandatory} & \textbf{Content Desc.} & \textbf{Scope} \\
\midrule
Video games & PEGI & 3, 7, 12, 16, 18 & Yes$^*$ & Yes (8 types) & Europe (38 countries) \\
Video games & ESRB & E, E10+, T, M, AO & Voluntary & Yes (30+ types) & USA, Canada \\
Film & MPA & G, PG, PG-13, R, NC-17 & De facto & Yes & USA \\
Film & BBFC & U, PG, 12A, 15, 18 & Yes & Yes & UK \\
Television & TV Parental Guidelines & TV-Y to TV-MA & Voluntary & Yes (5 types) & USA \\
Streaming video & AVMSD ratings & Varies by country & Yes & Varies & EU \\
\textbf{Music} & \textbf{RIAA PA} & \textbf{Binary (E/--)} & \textbf{Voluntary} & \textbf{No} & \textbf{USA} \\
\bottomrule
\multicolumn{6}{l}{\scriptsize $^*$Mandatory in most European countries through national legislation adopting PEGI.}
\end{tabular}
\end{table}

\subsection{Precedents in Music Content Regulation}

Although no country has implemented a comprehensive age-based rating system for music comparable to PEGI, several jurisdictions have taken regulatory steps that serve as partial precedents. South Korea's \textit{Korea Media Rating Board} (KMRB) classifies music videos and, in certain cases, audio tracks that are deemed inappropriate for minors, restricting their broadcast and online distribution to age-verified audiences \citep{KMRB2023}. Germany's \textit{Bundeszentrale f\"ur Kinder- und Jugendmedienschutz} (BzKJ), successor to the BPjM, maintains a list of media products---including music albums---deemed harmful to minors, restricting their advertising and distribution \citep{BzKJ2021}. In Australia, the \textit{Classification Board} has the authority to classify music with explicit content, and material classified as RC (Refused Classification) is effectively restricted from sale \citep{AustralianClassification2020}. The United Kingdom's \textit{Online Safety Act 2023} imposes duties on platforms to protect children from harmful content, which includes sexually explicit material in any medium, though it does not prescribe a specific music rating system \citep{UKOnlineSafety2023}.

These precedents, while fragmentary, demonstrate that the regulatory treatment of music content is not without basis and that there exists both legal foundation and political will for more systematic approaches. The critical missing element has been \textit{technological}: the absence of a scalable, reliable automated classification system that could make a granular rating system practically feasible for the vast catalogs of modern streaming platforms, which contain tens of millions of tracks. The present paper's technical contribution---a fine-tuned LLM capable of detecting sexually explicit content with high precision---addresses precisely this bottleneck.

\subsection{Proposed Framework: A Multi-Tier Rating System for Music}

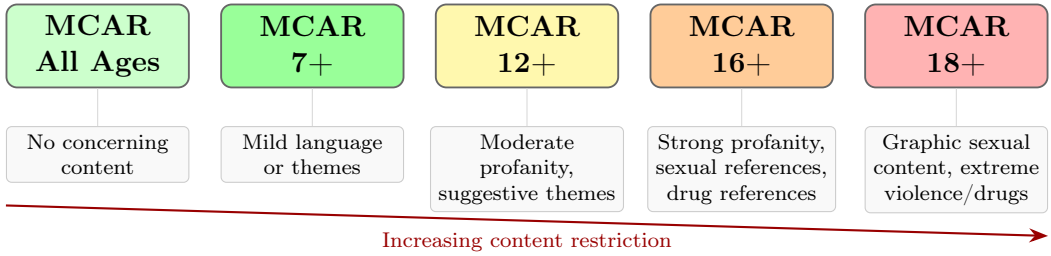
\begin{figure}[t]
\centering
\begin{tikzpicture}[
    tier/.style={rectangle, rounded corners=4pt, minimum width=2.4cm, minimum height=1.1cm,
        align=center, font=\small\bfseries, draw=black!60, line width=0.7pt},
    desc/.style={rectangle, rounded corners=2pt, draw=gray!40, fill=gray!5,
        minimum width=2.4cm, align=center, font=\scriptsize, inner sep=3pt},
    arrow/.style={-{Stealth[length=2.5mm]}, thick, draw=black!40}
]

\node[tier, fill=green!20] (t1) {MCAR\\All Ages};
\node[tier, fill=green!40, right=0.4cm of t1] (t2) {MCAR\\7+};
\node[tier, fill=yellow!40, right=0.4cm of t2] (t3) {MCAR\\12+};
\node[tier, fill=orange!40, right=0.4cm of t3] (t4) {MCAR\\16+};
\node[tier, fill=red!30, right=0.4cm of t4] (t5) {MCAR\\18+};

\node[desc, below=0.5cm of t1] (d1) {No concerning\\content};
\node[desc, below=0.5cm of t2] (d2) {Mild language\\or themes};
\node[desc, below=0.5cm of t3] (d3) {Moderate\\profanity,\\suggestive themes};
\node[desc, below=0.5cm of t4] (d4) {Strong profanity,\\sexual references,\\drug references};
\node[desc, below=0.5cm of t5] (d5) {Graphic sexual\\content, extreme\\violence/drugs};

\draw[gray!30, thin] (t1.south) -- (d1.north);
\draw[gray!30, thin] (t2.south) -- (d2.north);
\draw[gray!30, thin] (t3.south) -- (d3.north);
\draw[gray!30, thin] (t4.south) -- (d4.north);
\draw[gray!30, thin] (t5.south) -- (d5.north);

\draw[arrow, draw=red!60!black] ([yshift=-0.3cm]d1.south west) -- ([yshift=-0.3cm]d5.south east)
    node[midway, below, font=\scriptsize, text=red!60!black] {Increasing content restriction};

\end{tikzpicture}
\caption{Proposed Music Content Age Rating (MCAR) framework with five age-based tiers and associated content descriptors. The system is modeled on PEGI's proven multi-tier structure, adapted for the specific content dimensions relevant to music lyrics.}
\label{fig:mcar_tiers}
\end{figure}

Drawing on the PEGI model and informed by the technical capabilities demonstrated in this study, we propose a multi-tier age-based content rating system for music, which we term \textit{Music Content Age Rating} (MCAR). The proposed framework comprises five age tiers with associated content descriptors, as illustrated in Figure~\ref{fig:mcar_tiers}.

The operational implementation of MCAR would proceed in three stages. In the \textit{automated classification} stage, the lyrics of each track would be analyzed by a suite of fine-tuned LLMs---one specialized for each content dimension (sexual content, violence, drug references, profanity)---producing a vector of content scores. In the \textit{threshold mapping} stage, these scores would be mapped to age tiers through calibrated thresholds established by expert panels. In the \textit{human review} stage, tracks near classification boundaries or flagged by users would be reviewed by trained human moderators, ensuring that edge cases receive the contextual judgment that automated systems cannot fully replicate. This hybrid architecture balances the scalability of automated classification with the reliability of human oversight, following the best practices identified by \citet{Markov2023} for content moderation at scale.

\subsection{Policy Analysis: PESTEL Framework}

\begin{figure}[t]
\centering
\begin{tikzpicture}[
    factor/.style={rectangle, rounded corners=4pt, minimum width=2.6cm, minimum height=0.9cm,
        align=center, font=\small\bfseries, draw=black!60, line width=0.7pt},
    detail/.style={rectangle, rounded corners=2pt, draw=gray!40, fill=white,
        text width=5.5cm, align=left, font=\scriptsize, inner sep=4pt},
    connector/.style={draw=gray!50, thick}
]

\node[factor, fill=blue!20] (P) at (0, 0) {Political};
\node[factor, fill=green!20] (E) at (0, -1.5) {Economic};
\node[factor, fill=yellow!20] (S) at (0, -3.0) {Social};
\node[factor, fill=orange!20] (T) at (0, -4.5) {Technological};
\node[factor, fill=purple!15] (Ev) at (0, -6.0) {Environmental};
\node[factor, fill=red!15] (L) at (0, -7.5) {Legal};

\node[detail, right=1.0cm of P] (Pd) {Cross-party support for child protection; need for international harmonization (EU, LATAM); industry lobbying resistance};
\node[detail, right=1.0cm of E] (Ed) {Implementation costs for platforms (API integration); compliance costs for labels; potential revenue from ``safe'' playlists};
\node[detail, right=1.0cm of S] (Sd) {Strong public demand for child protection; cultural variation in ``explicit'' thresholds; generational attitudes toward regulation};
\node[detail, right=1.0cm of T] (Td) {LLM-based classification now viable (this study); streaming APIs enable enforcement; age verification technology maturing};
\node[detail, right=1.0cm of Ev] (Evd) {Minimal direct environmental impact; computational costs of large-scale classification; digital infrastructure requirements};
\node[detail, right=1.0cm of L] (Ld) {Free expression protections (1st Amend., Art.\ 10 ECHR); AVMSD precedent for video; GDPR constraints on age verification};

\draw[connector] (P.east) -- (Pd.west);
\draw[connector] (E.east) -- (Ed.west);
\draw[connector] (S.east) -- (Sd.west);
\draw[connector] (T.east) -- (Td.west);
\draw[connector] (Ev.east) -- (Evd.west);
\draw[connector] (L.east) -- (Ld.west);

\end{tikzpicture}
\caption{PESTEL analysis of the proposed Music Content Age Rating (MCAR) framework. Each macro-environmental dimension is assessed for factors that facilitate or constrain the implementation of an age-based music content rating system.}
\label{fig:pestel}
\end{figure}
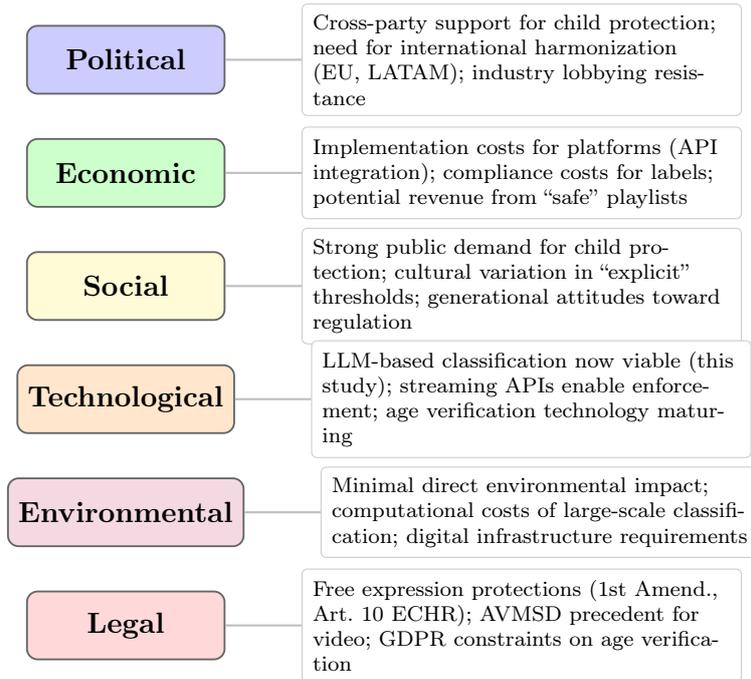

To assess the macro-environmental factors that would influence the adoption of MCAR, we employ the PESTEL framework, a widely used tool in public policy and strategic analysis that examines Political, Economic, Social, Technological, Environmental, and Legal dimensions \citep{Yuksel2012}. Figure~\ref{fig:pestel} presents the PESTEL analysis applied to the proposed music content rating system.

The PESTEL analysis reveals a broadly favorable environment for the adoption of MCAR. The \textit{political} dimension is characterized by strong cross-party support for child online safety measures in most democracies, as evidenced by the UK Online Safety Act 2023 and the EU's Digital Services Act, though implementation requires international coordination given the global nature of music streaming. The \textit{economic} dimension presents a mixed picture: while implementation imposes compliance costs on platforms and labels, it also creates commercial opportunities through differentiated ``family-safe'' product tiers and premium parental control features. The \textit{social} dimension is highly favorable, as public concern about children's exposure to explicit content is consistently high across cultures, though the specific thresholds of what constitutes ``explicit'' vary across communities and generations. The \textit{technological} dimension is newly favorable thanks to advances in LLM-based content classification---the present study being a direct contribution---which make automated, scalable rating technically feasible for the first time. The \textit{legal} dimension presents the most significant challenges, as any mandatory content rating system must navigate the tension between child protection and freedom of artistic expression, a tension that different legal traditions resolve differently.

\subsection{Kingdon's Multiple Streams Framework}

\begin{figure}[t]
\centering
\begin{tikzpicture}[
    stream/.style={rectangle, rounded corners=6pt, minimum width=3.5cm, minimum height=1.4cm,
        align=center, font=\small, draw=black!60, line width=0.8pt},
    window/.style={ellipse, minimum width=3.5cm, minimum height=1.6cm,
        align=center, font=\small\bfseries, draw=red!70!black, fill=red!10, line width=1pt},
    item/.style={font=\scriptsize, text width=3.2cm, align=left},
    arrow/.style={-{Stealth[length=3mm]}, thick, draw=black!50}
]

\node[stream, fill=blue!12] (prob) at (-4.5, 2) {\textbf{Problem Stream}};
\node[stream, fill=green!12] (pol) at (0, 2) {\textbf{Policy Stream}};
\node[stream, fill=orange!12] (polit) at (4.5, 2) {\textbf{Politics Stream}};

\node[item, below=0.3cm of prob] (pi1) {$\bullet$ Research on harm to minors \citep{Martino2006}};
\node[item, below=0.0cm of pi1] (pi2) {$\bullet$ Regulatory gap in music vs.\ other media};
\node[item, below=0.0cm of pi2] (pi3) {$\bullet$ High-profile incidents and public outcry};

\node[item, below=0.3cm of pol] (po1) {$\bullet$ LLM-based classification (this paper)};
\node[item, below=0.0cm of po1] (po2) {$\bullet$ PEGI model adaptable to music};
\node[item, below=0.0cm of po2] (po3) {$\bullet$ Hybrid human--AI review architectures};

\node[item, below=0.3cm of polit] (pt1) {$\bullet$ Online safety legislation momentum};
\node[item, below=0.0cm of pt1] (pt2) {$\bullet$ Cross-party child protection consensus};
\node[item, below=0.0cm of pt2] (pt3) {$\bullet$ Industry self-regulation fatigue};

\node[window] (win) at (0, -3.8) {\textbf{Policy Window}\\MCAR Adoption};

\draw[arrow] (pi3.south) .. controls (-4.5, -2.3) and (-2, -3.2) .. (win.west);
\draw[arrow] (po3.south) -- (win.north);
\draw[arrow] (pt3.south) .. controls (4.5, -2.3) and (2, -3.2) .. (win.east);

\end{tikzpicture}
\caption{Application of Kingdon's Multiple Streams Framework to the proposed Music Content Age Rating system. A policy window opens when the problem stream (documented harm, regulatory gap), the policy stream (technically feasible solutions), and the politics stream (legislative momentum, public demand) converge.}
\label{fig:kingdon}
\end{figure}
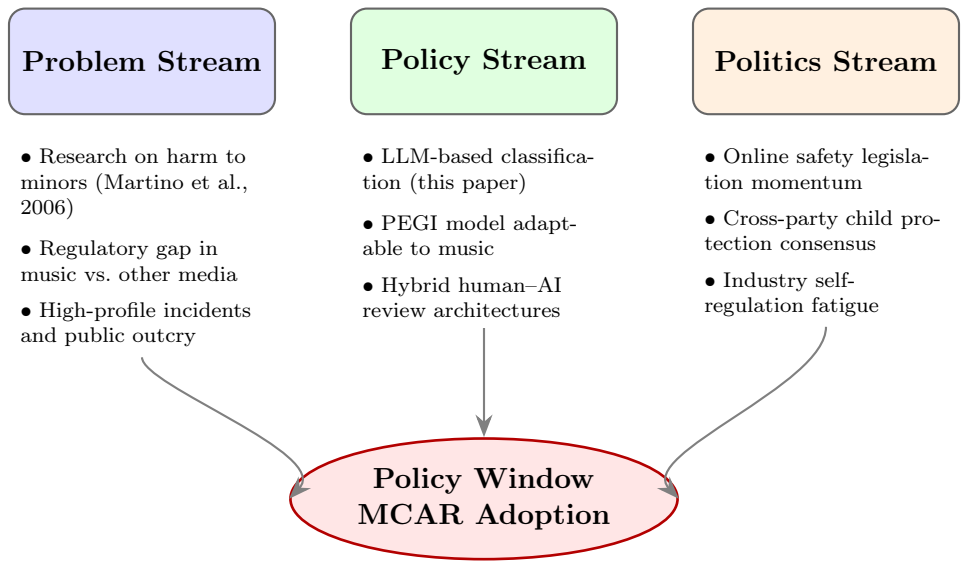

To analyze the political feasibility and timing of MCAR adoption, we apply Kingdon's Multiple Streams Framework (MSF), one of the most influential models in public policy theory for understanding how issues reach the governmental agenda \citep{Kingdon2011}. The MSF posits that policy change occurs when three largely independent streams---the \textit{problem stream}, the \textit{policy stream}, and the \textit{politics stream}---converge at a \textit{policy window}, creating an opportunity for policy entrepreneurs to advance their proposals. Figure~\ref{fig:kingdon} illustrates the application of this framework to the music content rating domain.

In the \textit{problem stream}, the issue of children's exposure to sexually explicit music lyrics has been documented by a growing body of research. \citet{Martino2006} established the association between exposure to degrading sexual content in music and adolescent sexual behavior, and \citet{DiezGutierrez2023} documented the reproduction of sexist stereotypes through reggaeton lyrics. The regulatory gap identified in Table~\ref{tab:rating_systems}---music as the only major entertainment medium without a granular rating system---constitutes a focusing event that draws attention to the problem.

In the \textit{policy stream}, the technical feasibility of automated lyric classification has been demonstrated by the present study and by prior work such as \citet{Rospocher2020} and \citet{Chen2023}. The existence of a proven institutional model (PEGI) that could be adapted to music provides a ready-made policy template, reducing the design costs and uncertainty associated with novel regulatory proposals. The hybrid human--AI architecture proposed in this paper addresses the reliability concerns that have historically prevented the adoption of fully automated content classification.

In the \textit{politics stream}, the current political environment in Europe and elsewhere is characterized by strong legislative momentum toward online child safety, exemplified by the UK Online Safety Act 2023, the EU Digital Services Act, and various national initiatives. There is broad cross-party consensus that platforms bear responsibility for protecting minors from harmful content, and the perceived failure of industry self-regulation---the RIAA's voluntary system being a prominent example---has increased political appetite for mandatory measures.

The convergence of these three streams suggests that a \textit{policy window} for the adoption of a music content rating system may be approaching. The role of \textit{policy entrepreneurs}---researchers, advocacy groups, and regulators who champion the proposal---will be critical in translating this convergence into concrete legislative action.

\subsection{Potential Objections and Counterarguments}

Any proposal to rate or classify music content by age will encounter significant objections, which must be addressed transparently to build a credible policy case.

\textbf{Freedom of expression and censorship.} The most fundamental objection is that age-based content rating constitutes a form of censorship that infringes on artistic freedom. This concern is legitimate but must be carefully distinguished from the actual mechanism proposed. Age rating does not prohibit, suppress, or alter content; it provides information to consumers and parents, enabling informed choice. The PEGI system has operated for over two decades without being characterized as censorship in European jurisprudence, and the same principle applies: an 18+ rating does not ban a song but informs listeners and enables parental controls. The European Court of Human Rights has consistently held that proportionate restrictions on minors' access to harmful content are compatible with Article~10 of the European Convention on Human Rights, provided they are prescribed by law and pursue a legitimate aim.

\textbf{Cultural relativism and subjectivity.} Critics may argue that the definition of ``sexually explicit'' is culturally contingent and cannot be standardized across jurisdictions. This is a valid concern, addressed in the MCAR framework through national or regional calibration of classification thresholds, analogous to how PEGI ratings are interpreted and enforced differently in practice across member countries. The use of expert panels representative of diverse cultural perspectives, combined with transparent classification criteria, would further mitigate this risk. The inherent subjectivity documented by \citet{Fell2019} and \citet{Darroch2014} does not invalidate the enterprise but rather underscores the need for human oversight alongside automated classification.

\textbf{Implementation costs and industry resistance.} The music industry may resist mandatory rating on grounds of cost and operational burden. However, the automated classification backbone of MCAR dramatically reduces the marginal cost per track compared to fully manual rating systems. Furthermore, major streaming platforms already invest heavily in content moderation infrastructure for other purposes (copyright enforcement, hate speech detection), and the incremental cost of integrating lyric classification into existing pipelines is modest in comparison. The economic analysis within the PESTEL framework suggests that compliance costs may be offset by new revenue opportunities in the ``family-safe'' market segment.

\textbf{Chilling effect on artists.} A related concern is that artists may self-censor to avoid restrictive ratings, thereby impoverishing musical creativity. Empirical evidence from the video game industry suggests that the PEGI and ESRB systems have not prevented the creation of mature-rated content; indeed, some of the industry's most commercially successful titles carry 18+/M ratings. A transparent, predictable rating system may actually benefit artists by providing clear expectations, reducing the arbitrary and inconsistent application of the current binary label, and channeling mature content toward its intended adult audience.

\textbf{Evasion and circumvention.} Finally, some may argue that age ratings are easily circumvented by tech-savvy minors. While no age-gating system is perfectly enforceable, the combination of platform-level parental controls, age verification at account creation, and content-aware recommendation filtering creates multiple layers of protection that, collectively, significantly reduce exposure. The objective is meaningful risk reduction, not perfect elimination---a standard that is accepted in every other domain of child safety regulation.

\section{Conclusions}
\label{sec:conclusions}

This paper has presented an approach to the automatic detection of sexually explicit content in Spanish-language reggaeton and trap lyrics through the fine-tuning of a Generative Pre-trained Transformer model on a curated, expert-annotated corpus of 100 songs. The experimental results demonstrate that the fine-tuned model achieves 87\% accuracy, 100\% precision, and 100\% specificity after a feedback-driven refinement loop, substantially outperforming a non-customized ChatGPT baseline and confirming the value of domain-specific adaptation for tasks that require cultural and linguistic sensitivity. The null hypothesis---that traditional text classification methods are as effective as transformer-based approaches for this task---has been rejected on the basis of the empirical evidence, and the alternative hypothesis---that fine-tuned transformer models achieve higher accuracy---has been confirmed.

The findings carry both technical and societal implications. Technically, they demonstrate that transfer learning enables high classification performance even with small training corpora, provided that the annotation is expert-driven and the fine-tuning procedure is carefully calibrated to the domain. Societally, they establish a viable framework for deploying automated content moderation tools on music streaming platforms, with direct applications in parental controls, content advisory labeling, and the protection of vulnerable audiences from exposure to sexually explicit lyrical material. The public policy analysis developed in Section~\ref{sec:policy} demonstrates that the conditions for adopting a multi-tier music content rating system---analogous to the PEGI system for video games---are increasingly favorable, as confirmed by both the PESTEL assessment and the convergence of Kingdon's three policy streams.

Several directions for future work emerge from the present study. The most immediate priority is the expansion of the training corpus across a wider range of artists, subgenres, languages, and cultural contexts, which would enhance the model's generalization capacity and reduce its dependence on the idiosyncrasies of the current dataset. The adoption of multi-annotator labeling protocols with inter-annotator agreement metrics would strengthen the reliability of the ground truth and enable the model to learn from the distribution of human judgments rather than from a single expert's perspective. Algorithmically, exploring lighter-weight model architectures such as DistilGPT or LLaMA could facilitate deployment on resource-constrained platforms, including mobile devices, without sacrificing classification performance. The integration of the model into a real-time content analysis pipeline, connected to the APIs of music streaming platforms such as Spotify or Apple Music, represents the logical next step toward translating the present research into a deployed system capable of analyzing songs at scale and in real time. Finally, a systematic investigation of the ethical implications of automated content classification in the creative arts---including questions of censorship, cultural bias, and the governance of algorithmic moderation---would provide a necessary complement to the technical contributions of this work.


\end{document}